\documentclass[eqsecnum,aps,showpacs]{revtex4}
\def\beq{\begin{equation}}
\def\eeq{\end{equation}}

\newcommand{\gsim}{
\raisebox{-0.8ex}{\mbox{$\stackrel{\textstyle >}{\sim}$}}}

\begin{document}

\title{
Tanaka-Tagoshi Parametrization of post-1PN\\ 
Spin-Free Gravitational Wave Chirps:\\
Equispaced  and Cardinal Interpolated Lattices\\
For First Generation Interferometric Antennas
}
\author{R.P. Croce and Th. Demma}
\affiliation{Wavesgroup, D.I.$^{3}$E., University of Salerno, Italy}
\vspace*{-.7cm}
\author{V. Pierro and I.M. Pinto} 
\affiliation{Wavesgroup, University of Sannio at Benevento, Italy}
\date{\today}

\begin{abstract}
The spin-free binary-inspiral parameter-space 
introduced by  Tanaka and Tagoshi  to  construct  
a uniformly-spaced  lattice of  templates  
at (and possibly beyond) $2.5PN$ order
is shown to work for all first generation interferometric 
gravitational wave antennas.
This allows to extend the minimum-redundant 
cardinal interpolation techniques of the correlator bank 
developed by the Authors 
to the highest available order PN  templates.
The total number of 2PN templates to be computed
for a minimal match $\Gamma=0.97$ is reduced
by a factor $\approx 4$, as in the 1PN case.
\end{abstract}


\maketitle

\section{Introduction}

Gravitational waves emitted by coalescing compact binaries
in their adiabatic inspiral phase have been accurately modeled \cite{chirps}
and are preferred candidates for the  direct detection 
of gravitational radiation of cosmic origin,
by the first generation 
of broadband laser interferometric  antennas, including 
TAMA300 \cite{TAMA}, GEO600 \cite{GEO}, 
the LIGOs \cite{LIGO} and VIRGO \cite{VIRGO}.

The best strategy for detecting signals of known form in colored gaussian 
stationary noise consists in correlating the detector output (data)
with a discrete family (lattice) of expected waveforms (templates), and using the
largest correlation as a detection statistic (maximum likelihood \cite{Helstrom}).
The above correlation is the (weighted) scalar product:
\beq
\langle a,g \rangle \equiv 2\left[
\int_{f_{i}}^{f_{s}} 
a(f) g^*(f)
\frac{df}{\Pi(f)} + C.C.
\right],
\label{eq:correl}
\eeq
where 
$(f_i, f_s)$ is the antenna spectral window,
$a(f)=h(f)+n(f)$  are the (noise-corrupted, spectral) data, 
$h(f)$ is a (possibly null) signal 
and $n(f)$ a realization of the antenna noise,
$g(f)$ is a template, 
$\Pi(f)$ is the (one-sided) antenna noise power spectral density 
and C.C. denotes the complex conjugate. 
The detection statistic has to be compared to a threshold, which
depends on the allowed false alarm probability.
At any fixed false-alarm probability, the largest probability of detection
is obtained iff a template is exactly {\em matched}  to the signal. 
The matching between  $h$  and  $g$  is measured by their {\em overlap}, 
\beq
{\cal O}(h,g)=\frac{\langle h,g \rangle}{||h||\!\cdot\!||g||} \leq 1,
\eeq
where the norm  $||u||=\langle u,u \rangle^{1/2}$.
The template lattice should be designed in such a way that for {\em any}
admissible signal, one can always find {\em at least one} template in the lattice 
such that \cite{lost_ev} the overlap is never less than a prescribed $\Gamma$. 

In the straightforward restricted \cite{restrict}  post-newtonian 
stationary-phase approximation \cite{stat_phas}
the spectral form of a null-eccentricity binary inspiral signal is:
\beq
h(f)=Af^{-7/6} \exp \left\{j\left[2\pi f T_c-\phi_c+\Psi^{(\nu)}(f,\vec{\xi})-\pi/4 \right]\right\},
\label{eq:signal}
\eeq
where $A$ is a (real, unknown) constant, depending on the source position, 
$T_c$ is the fiducial coalescency time \cite{fid_Tc},
$\phi_c$ is the phase at $t=T_c$, 
and  \cite{Blanchet}:
\beq
\Psi^{(\nu)}(f,\vec{\xi})=
\sum_{i=1}^{2\nu+1} \zeta_i(f) \theta^i(\vec{\xi}),
\label{eq:phase}
\eeq
where $\nu$ is the PN order,
and $\vec{\xi}$ represents the remaining (intrinsic) source parameters.
The functions $\zeta_i$ and $\theta^i$  
are presently known up to 2.5PN order \cite{3PN_stall} 
and are collected in Table-I \cite{notation}
in terms of the intrinsic parameters 
$m$ (total mass of the binary), 
$\eta=\mu/m$  ($\mu$ being the reduced mass),
$\beta$ and $\sigma$ (spin-orbit and spin-spin terms, respectively \cite{Apo96}).

Maximizing explicitly the overlap over 
the (irrelevant, {\em extrinsic}) parameters $\Delta\phi_c$ and $\Delta T_c$, 
one can write the template-lattice design prescription as:
\beq
\forall h \in {\cal S},~\exists g \in {\cal L}~:~
M(h,g) \equiv
\sup_{\Delta T_c}~
\frac{
\displaystyle{
\left| 
\int_{f_{i}}^{f_{s}} 
f^{-7/3}
\exp\left\{j\left[2\pi f \Delta T_c+\Delta\Psi^{(\nu)}(f,\vec{\xi}_h,\vec{\xi}_g) \right]\right\}
\frac{df}{\Pi(f)}
\right|
}
}{
\displaystyle{
\int_{f_{i}}^{f_{s}} 
f^{-7/3}
\frac{df}{\Pi(f)}
}
} \geq \Gamma
\label{eq:match}
\eeq
where the partially maximised overlap $M(f,g)$ is called  the {\em match},
${\cal S}$ and ${\cal L}$ are the signal-space and template-lattice, respectively,
and:
\beq
\Delta\Psi^{(\nu)}(f;\vec{\xi}_h,\vec{\xi}_g)=
\Psi^{(\nu)}(f,\vec{\xi}_h)-
\Psi^{(\nu)}(f,\vec{\xi}_g)=
\sum_{i=1}^{2\nu+1} \zeta_i(f) 
\left[
\theta^i(\vec{\xi}_h)-
\theta^i(\vec{\xi}_g)
\right].
\label{eq:delta_psi}
\eeq

Whenever the match is only a function 
of the signal/template parameter {\em differences},
the template lattice is {\em uniform}, 
i.e., the lattice spacing is {\em constant}
throughout the template parameter space, 
and depends uniquely on $\Gamma$ \cite{OweSat}.
This is obviously the case
iff the function $\Delta\Psi^{(\nu)}$ depends, in turn,
on the signal/template parameter differences only.

Uniformity is a key property for efficient
implementation of the lattice.
In a uniform lattice, templates corresponding to adjacent nodes
differ only by a {\em fixed} phase factor, which makes the
template family construction computationally inexpensive.
Furthermore, 
in view of the quasi band-limited nature of the match
as a function of the signal/template parameter differences,
cardinal interpolation among the correlators 
can be used as shown in \cite{card0}, \cite{card1}
to substantially reduce the number of templates 
required for a prescribed $\Gamma$.

The 1D newtonian ($\nu=0$) and 2D first-order post-newtonian ($\nu=1$) 
waveforms can be easily parametrized 
so as to make the phase-difference (\ref{eq:delta_psi}) a function 
of the signal/template parameter {\em differences} only \cite{how}.
On the other hand, according to present understanding,
in order to keep $\Gamma \agt 0.97$
when working with {\em true} data, 
$2PN$  (or higher) templates will be needed,
although spin-free ones should be adequate \cite{which_tmp}.

In the following  we shall refer to 2PN templates. 
These are the best available ones for the purpose of detection/estimation.
Indeed, as shown in \cite{DamIyeSat},  2.5PN templates 
yield generally  {\em poorer} overlaps (and larger  biases)
with (exact) numerically generated waveforms
as compared to 2PN, as an effect of the peculiar
(oscillating) behaviour  of PN-approximant sequences.

Unfortunately,  at  PN orders  $\nu > 1$, as shown in Sect. II,
it is {\em strictly  impossible} to parametrize the  spin-free  waveforms
so as to make the phase-difference (\ref{eq:delta_psi})  a function 
of the signal/template parameter {\em differences} only.
As a result, the choice of template placement and spacing
becomes, in principle, rather unwieldy, 
and the cardinal interpolation techniques
discussed in \cite{card0}, \cite{card1} cannot be applied in any
straightforward fashion.

Recently, Tanaka and Tagoshi \cite{TaTa} suggested
a clever way to circumvent this difficulty,
using a simple and elegant  geometric argument.
The scope of this paper is to illustrate the effectiveness of the
Tanaka-Tagoshi construction 
for {\em all}  first generation interferometers 
(TAMA300, GEO600, LIGO-I and VIRGO)
under a given minimal match constraint (see Section III).
In the following we use geometrized units ($G=c=1$) throughout.

\section{The Intrinsic Curvature of The Spin-Free Parameter Space Manifold}

The Tanaka-Tagoshi construction is best understood
in the geometric language first introduced in \cite{Owe96}.
The match between $h$ and $g$ can be accordingly written
\cite{parabolic}:
\beq
M(h,g) = 1 - 
G_{rs} \Delta\theta^{r}\Delta\theta^{s} + \dots,
\label{eq:match1}
\eeq
where $\Delta \theta^i = \theta^i(\vec{\xi}_h)-\theta^i(\vec{\xi}_g)$, 
and the ($2\nu+1$)-dimensional
metric $G_{rs}$ is defined in Appendix A. 

Let $m_{min} \leq m_1 \leq m_2 \leq m_{max}$
the spin-free waveform parameter space,
$m_1$, $m_2$ being the companion masses \cite{invar}.
In the $(2\nu\!+\!1)$-dimensional PN  space $\theta^i$
the parameter space is mapped into a three-vertex 2D manifold ${\cal P}$.
Exploiting the dependence of  $\theta^{i}$
on $m_1$ and $m_2$ one can rewrite  (\ref{eq:match1}) as:
\beq
M(h,g)=1-ds^2,~~~ds^2=g_{pq}\Delta m^{q} \Delta m^{p},
\eeq
where
\beq
\vspace*{-2mm}
g_{pq}=G_{rs}
\frac{\partial \theta^{r}}{\partial m_p}
\frac{\partial \theta^{s}}{\partial m_q},
\label{eq:2Dmetric}
\eeq
is a 2D Finsler metric \cite{Spivak} on ${\cal P}$.

In order to find a chirp waveform parametrization
which makes the match a function of the
source-template {\em parameter differences} only,
one should be able to find a coordinate transformation
$(m_1,m_2) \rightarrow (x_1,x_2)$  such that:
\beq
M(h,g)=1-\delta_{pq} \Delta x^{p} \Delta x^{q},
\label{eq:flatness}
\eeq
where $\delta_{pq}$ is the 2D euclidean metric.
As anticipated, the required coordinate transformation does {\em not} exist, 
in general. 
This is due to the fact that the post-1PN manifold ${\cal P}$ is {\em not} globally flat,
as can be seen synthetically from its Gaussian curvature $K$ \cite{GaussK}.
The gaussian curvature of   ${\cal P}$  as a function of  $m_1,m_2$
in  $0.2 M_{\odot} \leq m_{1,2}  \leq 10M_{\odot}$.
is shown in Fig. 1, for the special case of LIGO-I  at 2PN order.
The absolute curvature is maximum for $m_1\!=\!m_2$.
This property turns out to be common to {\em all}  
first-generation  antennas, whose relevant parameters
have been collected  in Table-II. 
In Fig. 2  the curvature of  ${\cal P}$ is displayed as a function of
$m_1\!=\!m_2$ (worst case) 
for TAMA300,  GEO600, LIGO-I and VIRGO  at 2PN order. 

\section{The Tanaka-Tagoshi Coordinates}

Even though ${\cal P}$ is {\em not}  flat, 
following Tanaka and Tagoshi  it is still possible
to place the templates on 
a {\em flat} manifold chosen 
{\em close} to ${\cal P}$ in a suitable sense.
The Tanaka-Tagoshi construction 
can be simply phrased as follows:
i) introduce a linear coordinate transformation
$\vec{\theta} \rightarrow \vec{x}$  
such that {\em all  three} vertices of  the manifold ${\cal P}$
are  brought onto the $(x_1,x_2)$ plane 
of the (orthogonal) coordinate system $x_i$
(see Appendix-B for details), 
and then 
ii)  take  the resulting three-vertex {\em flat 2D simplex}   
${\cal T}$  in the $(x_1,x_2)$ plane
as the signal/template parameter space.

The flat simplexes ${\cal T}$  corresponding to
$0.2 M_{\odot} \leq m_1 \leq m_2  \leq 10M_{\odot}$
are shown in Fig. 3
for TAMA300, GEO600, LIGO-I and VIRGO  at 2PN order.
Their  measures (areas)  are collected in Table-III.

Enforcing a minimal-match constraint,
in the $(x_1,x_2)$  plane yields 
the {\em uniform} square-mesh lattice sidelength
\beq
\Delta=\left[
2(1-\Gamma)
\right]^{1/2}.
\label{eq:sidelenght}
\eeq
The simplex areas divided by $\Delta^2$  
(area spanned by each template at  a given $\Gamma$)
provide close estimates of the corresponding total number
of templates required, also listed in Table-III for $\Gamma=0.97$,
for TAMA300, GEO600, LIGO-I and VIRGO at  2PN  order.

\subsection{The Minimal Match Error}

Pictorial representations of the departure of  ${\cal P}$ from ${\cal T}$
are shown  Fig.  4,  where the simplex ${\cal T}$
corresponding to  $0.2 M_{\odot} \leq m_1 \leq m_2  \leq 10M_{\odot}$
is displayed  together with its (euclidean) distance 
$\delta=(x_3^2+x_4^2+x^2_5)^{1/2}$ 
from ${\cal P}$  as a function of $x_1, x_2$, 
for TAMA300, GEO600, LIGO-I and VIRGO  at 2PN order.
The obvious question is to what extent 
does the shown difference spoil 
the minimal-match condition enforced in ${\cal T}$,
due to  the fact that 
the {\em true} waveform space is ${\cal P}$.

Let  $h$  a  ($\nu-$PN)  chirp signal,  and let
$H\equiv(x_1^{(h)}, x_2^{(h)},\dots, x_{2\nu+1}^{(h)})$
and
$\tilde{H} \equiv  (x_1^{(h)}, x_2^{(h)},0,\dots,0)$
the corresponding points  in  ${\cal P}$ and ${\cal T}$.
By construction, 
$\forall~h$ in the admissible spin-free
source parameter range  
$m_{min} \leq m_1 \leq m_2 \leq m_{max}$,
there will be at least one lattice node 
$\tilde G \equiv (x_1^{(g)}, x_2^{(g)},0,\dots,0) \in {\cal T}$
such that:
\beq
M(\tilde{h},\tilde{g}) = 1-\delta_{pq}
(x_p^{(\tilde h)}\!-\!x_p^{(\tilde g)})
(x_q^{(\tilde h)}\!-\!x_q^{(\tilde g)}) = \Gamma,
\eeq
where $\tilde h$ and $\tilde g$ are the waveforms
corresponding to  $\tilde H$, $\tilde G$, respectively.
On the other hand, from (\ref{eq:match1}), (\ref{eq:2Dmetric}) 
\beq
M(h,g)=1-\gamma_{pq}
(x_p^{(h)}\!-\!x_p^{(g)})
(x_q^{(h)}\!-\!x_q^{(g)})=
1-\gamma_{pq}
(x_p^{(\tilde h)}\!-\!x_p^{(\tilde g)})
(x_q^{(\tilde h)}\!-\!x_q^{(\tilde g)})
\eeq
where $p,q=1,2$, and \cite{detail}:
\beq
\gamma_{pq}=G_{rs}
\frac{\partial \theta^r}{\partial x^p}
\frac{\partial \theta^s}{\partial x^q}
\label{eq:gamma}
\eeq
where $r,s=1,2,\dots,2\nu+1$.
Let $\vec{x} \rightarrow \vec{y}$ 
the (unit) rotation which diagonalizes the matrix
$\gamma_{pq}-\delta_{pq}$. Then:
$$
\left| 
M(h,g)\!-\!M(\tilde{h},\tilde{g})
\right|
=
\left|
\eta_1
(y_1^{(\tilde h)}\!-\!y_1^{(\tilde g)})^2\!+\!
\eta_2
(y_2^{(\tilde h)}\!-\!y_2^{(\tilde g)})^2
\right|
\leq
$$
\beq
\leq
\max\left( |\eta_1|,|\eta_2| \right)
\left[
(y_1^{(\tilde h)}\!-\!y_1^{(\tilde g)})^2+
(y_2^{(\tilde h)}\!-\!y_2^{(\tilde g)})^2
\right]=
\max\left( |\eta_1|,|\eta_2| \right)
(1\!-\!\Gamma)
\eeq
where $\eta_1,\eta_2$ are the eigenvalues of
the matrix $\gamma_{pq}-\delta_{pq}$.
The $\Gamma$-independent quantity:
\beq
\eta=
\frac{\left|
M(h,g)-M(\tilde{h},\tilde{g})
\right|
}{
1-\Gamma}=
\max\left( |\eta_1|,|\eta_2| \right)
\label{eq:bound}
\eeq
is a measure of the match degradation due to
using a {\em flattened} parameter space,
and is displayed in Fig.  5
as a function of $m_1,~m_2$
in  $0.2 M_{\odot} \leq m_{1,2} \leq 10 M_{\odot}$,
for the special case of LIGO-I at  2PN order.
The maximum value is attained for $m_1\!=\!m_2$. 
This property is common to {\em all}  
first-generation  antennas quoted  in Table-I. 
In Fig. 6  the quantity $\eta$ is displayed
as a function of  $m_1\!=\!m_2$ (worst case) 
for TAMA300, GEO600, LIGO-I and VIRGO at 2PN order.

It is seen that the minimal-match condition enforced 
using ${\cal T}$ as the waveform/template parameter space
is {\em not} spoiled significantly when
the signal (and the template) belongs to ${\cal P}$.

The above argument might perhaps be loosey at
{\em low}  values of $\Gamma$, as e.g. typically required  
for the initial steps of hierarchical search procedures \cite{Mohanty}, 
where the quadratic approximation (\ref{eq:match1}) 
might be no longer accurate enough \cite{parabolic}.
In order to evaluate the (maximum) minimal-match error under these
broader conditions, we generated  $10^4$ random pairs $(m_1,m_2)$, 
uniformly distributed in the allowed source parameter range
$0.2 M_{\odot} \leq m_1 \leq m_2 \leq 10M_{\odot}$.
For each of the above, we drew  a  closed-curve 
${\cal C} \subset {\cal T}$ centered around 
the corresponding point  $\tilde{F} \in {\cal T}$,
whose points $\tilde G$  correspond  to (all)  templates $\tilde{g}$  for which  
$M(\tilde{h},\tilde{g})=\Gamma$, and computed
\beq
\epsilon=
\sup_{\tilde{G} \in {\cal C}} 
\frac{
\left|
M(h,g)-M(\tilde{h},\tilde{g})
\right|
}{\Gamma}=
\sup_{\tilde{G} \in {\cal C}} 
\left[
1-\frac{M(h,g)}{\Gamma}
\right].
\label{eq:epsilon}
\eeq
The cumulative distributions of $\epsilon$
corresponding  to $\Gamma= 0.7, 0.8, 0.9$, 
for the special case of LIGO-I   at  2PN order,
are shown in Fig. 7.
Not unexpectedly, the Tanaka-Tagoshi coordinates are seen to work
also at relatively low values of $\Gamma$.

\subsection{Cardinal Interpolation Beyond 1PN}

In the obtained uniform lattice, the application
of the cardinal interpolation technique 
introduced in \cite{card0}  is straightforward.
For 1PN templates this latter yields
a fourfold reduction in the template density
and total number  at $\Gamma=.97$, 
as shown in \cite{card1}.
A main motivation of this study
has been to check whether a comparable reduction
in the number of templates could be obtained 
using  Tanaka-Tagoshi  parametrized post-1PN templates.
Numerical simulations show that this is indeed the case:
the 2PN  template density reduction as a function
of the minimal match $\Gamma$ is shown in Fig.  8.

\subsection{Computation of Templates}

In the following we shall denote the  (linear)
direct  ($\vec{\theta} \rightarrow \vec{x}$)
and inverse ($\vec{x} \rightarrow \vec{\theta}$)
Tagoshi-Tanaka transformation operators 
as $\Xi_{ij}$ and $\Theta^{ij}$, respectively.

The template corresponding to
the  node  $(x_{1,k},x_{2,k})$  of the (uniform) lattice 
in the $(x_1,x_2)$ plane
should be computed, in principle, 
by solving (e.g., numerically) the system:
\beq
\left\{
\begin{array}{l}
\Xi_{1i}\theta^i(m_{1k},m_{2k})=x_{1,k}\\
\Xi_{2i}\theta^i(m_{1k},m_{2k})=x_{2,k},\\
\end{array}
\label{eq:sys}
\right.
\eeq
to obtain the corresponding values  $m_{1k},m_{2k}$
of $m_1,m_2$, whereby
{\em all} the $\theta^i$ functions  and hence the ({\em exact})
template phase (\ref{eq:phase}) can be computed.
The above {\em exact} (and computationally  expensive)
procedure for computing the template
phases needs {\em  not} to be applied for {\em all}   lattice nodes. 
Indeed, as noted in  \cite{TaTa}, 
the {\em change} in the $\theta^i$ between the templates
at $(x_1,x_2)$ and $(x_1+\delta x_1,x_2+\delta x_2)$
is well approximated by:
\beq
\Delta\theta^i\approx\Theta^{i1}\delta x_1+\Theta^{i2}\delta x_2
\label{eq:approx_phas}
\eeq
as a consequence of the fact that $x_3,x_4,\dots,x_{2\nu+1}$
are {\em almost} constant throughout 
relatively {\em large} portions of the $(x_1,x_2)$ plane.
As a result, the   phase {\em difference} 
between neighbouring templates can be taken as  {\em uniform} 
throughout  {\em extended} regions of the $(x_1,x_2)$ plane,
which makes the template set construction relatively cheap.
One can accordingly compute the exact template phases
using Eq. (\ref{eq:sys}) for a set of {\em sparse} lattice nodes.
The phase of  templates belonging to neighbourhoods
of these sparse nodes can be obtained 
using  Eq. (\ref{eq:approx_phas}).
The {\em size} of the above neighbourhoods 
(i.e., equivalently, the distance between the sparse nodes)
should be chosen so as not to spoil 
the prescribed minimal match condition  \cite{trial}.

A possible natural strategy is to place the exact templates 
nearby  the   median-lines  of  ${\cal T}$, starting from their
common crossing point \cite{medians}. 
We checked by numerical simulation that a few 
(e.g., $\sim 50$ for LIGO-I at $\Gamma=0.97$)
exactly computed sparse templates
are sufficient to compute
all the remaining ones via (\ref{eq:approx_phas}),
without spoiling the minimal match condition 
throughout the whole waveform parameter range. 

\section{Conclusions}

~~~~~It has been shown that the Tanaka-Tagoshi coordinates
are effective to set up a uniform lattice of  2PN (and possibly higher)  
spin-free binary-inspiral templates 
under a given minimal match constraint,
for all 1st generation interferometric antennas.
This allows to extend readily
the minimum redundant cardinal interpolation techniques 
of the correlator bank introduced by the Authors
to the highest  available  order  PN templates,
yielding a computational gain comparable to the 1PN case.

Several possible variations of the 
Tanaka-Tagoshi method can be envisaged.
For instance,  a  2D  {\em simplex}  is used here
to approximate the parameter manifold. 
Using  a 2D {\em complex} could possibly  improve the accuracy.
Further,  the 2D simplex is obtained  by point-collocation, 
i.e. by enforcing contact between the vertices of the simplex ${\cal T}$  
and those of the manifold ${\cal P}$.
Other approximation philosophies could be used, 
including e.g., least mean squares, etc.

\section*{Acknowledgements}

~~~~~This work has been sponsored in part by the EC
through a senior visiting scientist grant to I.M. Pinto at NAO, Tokyo, JP.
I.M. Pinto wishes to thank the TAMA staff at NAO, and in particular prof. Fujimoto
Masa-Katsu and prof. Kawamura Seiji for gracious hospitality and stimulating discussions.
The authors also thank  prof.   B.S. Sathyaprakash for interesting  highlights,
and  prof.s  Tagoshi Hideyuki  and   Tanaka Takahiro  for helpful comments.


\section*{Appendix A  -  The PN Waveform Parameter Manifold}
~~~In this Appendix we rephrase the well known construction
of a metric in the binary-inspiral waveform parameter-space,
first introduced by Owen \cite{Owe96} 
and adopted by  Tanaka and Tagoshi \cite{TaTa},
using a slightly different  notation.
Define:
$$
u(h,g)=
\frac{
\displaystyle{
\int_{f_{i}}^{f_{s}}\!
f^{-7/3}
\exp\left\{j\left[2\pi f \Delta T_c
\!-\!\Delta \phi_c
\!+\!\Delta\Psi^{(\nu)}(f;\vec{\xi}_h,\vec{\xi}_g) \right]\right\}
\frac{df}{\Pi(f)}}
}{\displaystyle{
\int_{f_{i}}^{f_{s}}\!
f^{-7/3}
\frac{df}{\Pi(f)}
}},
\eqno{(A1)}
$$
and let:
$$
\langle
v(f)
\rangle =
\frac{
\displaystyle{\int_{f_i}^{f_s} df \frac{f^{-7/3}}{\Pi(f)} v(f) }
}{
\displaystyle{\int_{f_i}^{f_s} df \frac{f^{-7/3}}{\Pi(f)}  }
}.
\eqno{(A2)}
$$
It is readily proved that:
$$
u(h,g)=1
+j \langle \zeta_n \rangle \Delta\theta^n 
- \frac{1}{2} \langle \zeta_m \zeta_n  \rangle \Delta\theta^m  \Delta\theta^n+\dots,
\eqno{(A3)}
$$
where $m,n=0,1,\dots,2\nu+1$, and
$$
\zeta_0 = 2\pi (f/f_0),~~\theta_0 = f_0 T_c.
\eqno{(A4)}
$$
From equations $(A1)-(A4)$ one gets:
$$
\max_{\Delta\phi_c} 
\langle h,g \rangle = \left| u(h,g) \right|=
1-G'_{mn} \Delta\theta^m  \Delta\theta^n,
\eqno{(A5)}
$$
where $m,n=0,1,\dots,2\nu+1$ and
$$
G'_{mn} =\frac{1}{2}\left(
 \langle \zeta_m \zeta_n \rangle -
\langle \zeta_m \rangle  \langle \zeta_n \rangle
\right).
\eqno{(A6)}
$$
Equations $(A5)$ and $(A6)$  allow to express the match as follows:
$$
M(h,g) = \max_{\Delta T_c}
\max_{\Delta\phi_c}
\langle h,g \rangle =
\max_{\Delta T_c} \left| u(h,g) \right| =
1 - G_{pq}\Delta\theta^p\Delta\theta^q,
\eqno{(A7)}
$$
where now  $p,q=1,2,\dots,2\nu+1$ and  \cite{project}
$$
G_{pq} = G'_{pq} - \frac{G'_{p0}G'_{q0}}{G'_{00}}.
\eqno{(A8)}
$$

\section*{Appendix B - Constructing the Tanaka-Tagoshi Coordinates}
~~~~~Let  $m_{min} \leq m_1 \leq m_2 \leq m_{max}$
the allowed mass range, and let
$\vec{p}_i,~i=1,2,3$  the vertices
of the parameter space manifold  ${\cal P}$
corresponding to 
($m_1\!=\!m_2\!=\!m_{min}$), 
($m_1\!=\!m_{min}, m_2\!=\!m_{max}$) and
($m_1\!=\!m_2\!=\!m_{max}$), respectively,
in the  ($2\nu+1$)-dimensional  PN  parameter space with metric $G_{pq}$.
The Tanaka-Tagoshi coordinates are obtained as follows.
One first performs a (linear) coordinate transformation
$$
\vec{\theta}\prime=\mathbf{\Lambda}^{1/2}\mathbf{P} \vec{\theta}
\eqno{(B1)}
$$
to make the new coordinates orthogonal.
The required transformation is obtained from 
the Jordan decomposition of  $\mathbf{G}$,
$$
\mathbf{G} = \mathbf{P}^T \mathbf{\Lambda} \mathbf{P},
\eqno{(B2)}
$$
i.e., in component notation:
$$
G_{pq}=
P_p^{(m)}
\Lambda_{mn}
P_q^{(n)},
\eqno{(B3)}
$$
where $\Lambda_{mn}=\lambda^{(n)}\delta_{mn}$,
$\lambda^{(n)}$ being  the eigenvalues  and
$P_q^{(n)}$ the eigenvectors of  $\mathbf G$.

Next, one seeks a rotation $\vec{x} = {\mathbf Q} \vec{\theta}\prime$
which bears (modulo a trivial translation)  
{\em all}  vertices $\vec{p}~\prime_i,~i=1,2,3$  of the manifold ${\cal P}$
onto the coordinate plane $(x^1,x^2)$.
The problem is thus reduced to that of finding 
an orthogonal matrix $\mathbf Q$ such that:
$$
\left\{
\begin{array}{l}
\mathbf{Q}
(\vec{p}\prime_3-\vec{p}\prime_1)=\alpha_{11}\hat{x}_1\\
\mathbf{Q}
(\vec{p}\prime_2-\vec{p}\prime_1)=\alpha_{21}\hat{x}_1+\alpha_{22}\hat{x}_2,
\end{array}
\right.
\eqno{(B4)}
$$
where $\alpha_{11},~\alpha_{21},~\alpha_{22}$ are suitable
real numbers, and $\hat{x}_i$ denote the  (new, unit) basis vectors.
The system $(B4)$ does {\em not} have an unique solution for $\mathbf{Q}$. 
However a possible solution is readily obtained as follows.
Let $\mathbf{Z}$ the matrix constructed out of the (column) vectors: 
$$
\vec{p}\prime_3-\vec{p}\prime_1,~
\vec{p}\prime_2-\vec{p}\prime_1,~
\hat{\theta}\prime_3,~\dots, \hat{\theta}\prime_{2\nu+1}
\eqno{(B5)}
$$
It is readily shown that $\mathbf{Z}$ is non-singular.
The straightforward  (unique) QR  decomposition  of $\mathbf{Z}$, viz. \cite{QR} :
$$
\mathbf{Z}=\mathbf{Q}^{T} \mathbf{R},
\eqno{(B6)}
$$
where $\mathbf{R}$ is a lower triangular matrix,
is obviously a solution of:
$$
\mathbf{Q Z} = \mathbf{R}.
\eqno{(B7)}
$$
It is seen that the solution of  eq. $(B7)$   is  {\em also}
a solution of $(B4)$, and hence the sought $\mathbf{Q}$ in $(B4)$ is
the same as $\mathbf{Q}$ in $(B6)$.

\newpage
$$~$$
\begin{center}
\begin{tabular}{|c|c|c|c|}
\hline\hline
PN order & $i$ & $\zeta_i(f)$ & $\theta^i$ \\
\hline\hline
$0$ & $1$ &  
$\displaystyle{\left(\frac{f}{f_0}\right)^{-5/3}}$ & 
$\displaystyle{\frac{3}{128\eta}\left(\pi m f_0\right)^{-5/3}}$\\ 
\hline 
$0.5$ & $2$ &  
$\displaystyle{\left(\frac{f}{f_0}\right)^{-4/3}}$ & 
$\displaystyle{\frac{}{}0}$\\
\hline 
$1$ & $3$ &  
$\displaystyle{\left(\frac{f}{f_0}\right)^{-1}}$ &
$\displaystyle{\frac{5}{96\eta}\left(\frac{743}{336}+\frac{11}{4}\eta\right)
\left(\pi m f_0\right)^{-1}}$\\ 
\hline 
$1.5$ & $4$ &  
$\displaystyle{\left(\frac{f}{f_0}\right)^{-2/3}}$ &		 
$\displaystyle{\frac{3}{32\eta}\left(\frac{}{}\beta-4\pi\right)
\left(\pi m f_0\right)^{-2/3}}$\\ 
\hline
$2$ & $5$ &  
$\displaystyle{\left(\frac{f}{f_0}\right)^{-1/3}}$ &		 
$\displaystyle{\frac{15}{64\eta}\left(
\frac{3058673}{1016064}+\frac{5429}{1008}\eta+\frac{617}{144}\eta^2-\sigma\right)
\left(\pi m f_0\right)^{-1/3}}$\\ 
\hline
$2.5$ & $6$ &
$\displaystyle{\log\left(\frac{f}{f_0}\right)}$ &
$\displaystyle{\frac{\pi}{128 \eta}\left(\frac{38645}{252} + 5 \eta\right)}$\\
\hline\hline
\end{tabular}
$$~$$
Table I - Relevant to equation (\ref{eq:phase})
\end{center}
$$~$$
\begin{center}
\begin{tabular}{|c|c|c|c|c|c|}
\hline\hline
Antenna & $\Pi(f)$ & $\Pi_0~[Hz^{-1}]$ & $f_0~[Hz]$ & $f_{i}~[Hz]$ & $f_{s}~[Hz]$\\
\hline\hline
TAMA300 
& $\frac{\Pi_0}{32}\left\{\left(\frac{f_0}{f}\right)^5+13\left(\frac{f_0}{f}\right)+9\left[
1+\left(\frac{f}{f_0}\right)^2\right]\right\}$ 
& $2.4 \cdot 10^{-44}$ & $400$ & $75$ & $3400$\\ 
\hline
GEO600 & 
$\frac{\Pi_0}{5}\left[4\left(\frac{f_0}{f}\right)^{3/2}-2+3\left(\frac{f}{f_0}\right)^2\right]$ 
& $6.6 \cdot 10^{-45}$ & $210$ & $40$ & $1450$\\ 
\hline
LIGO-I & $\frac{\Pi_0}{3}\left[\left(\frac{f_0}{f}\right)^4+2\left(\frac{f}{f_0}\right)^2\right]$ 
& $4.4 \cdot 10^{-46}$ & $175$ & $40$ & $1300$\\ 
\hline
VIRGO & $\frac{\Pi_0}{4}\left[290\left(\frac{f_i}{f}\right)^5+2\left(\frac{f_0}{f}\right)+1+\left(\frac{f}{f_0}\right)^2\right]$ 
& $1.1 \cdot 10^{-45}$ & $475$ & $16$ & $2750$\\ 
\hline\hline
\end{tabular}
$$~$$
Table - II - Spectral windows and noise power spectral densities\\ 
of first generation interferometric GW antennas \cite{OweSat}.
\end{center}
$$~$$
\begin{center}
\begin{tabular}{|c|c|c|}
\hline\hline
Antenna &  Simplex Area $[sec^2]$& No. of  Templates at  $\Gamma=0.97$\\
\hline
TAMA300  &  $5987$   & $9.98\times 10^4$\\                                                                                                                                                                                                                                                                                                                                           
\hline
GEO600  & $38931$ & $6.49 \times 10^5$\\
\hline
LIGO-I  & $16842$  &  $2.81 \times 10^5$\\
\hline
VIRGO  &   $546689$ &  $9.11 \times 10^6$\\                                                                                                                                                   
\hline\hline
\end{tabular}
$$~$$
Table III - 2PN flat simplex areas and  number of templates at $\Gamma=.97$,\\ 
for $0.2 M_{\odot} \leq m_1 \leq m_2 \leq 10 M_{\odot}$.
\end{center}

\newpage
\begin{center}
{\Large \bf CAPTIONS TO THE FIGURES}
\end{center}
\normalsize
$$~$$
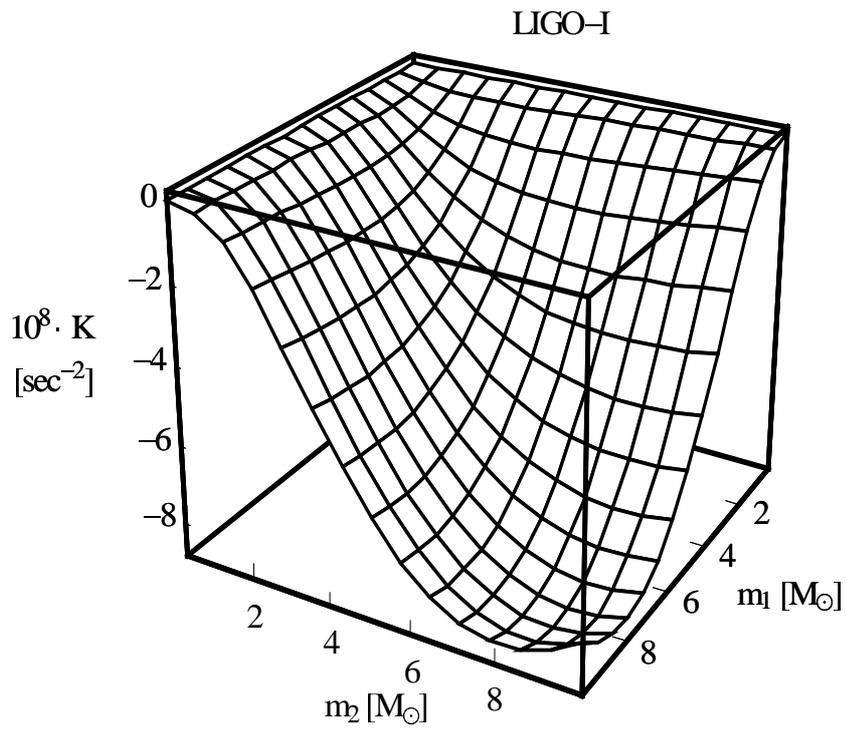
Fig. 1 - The gaussian curvature $K$ of ${\cal P}$ vs. $m_1$, $m_2$ for LIGO-I.
$$~$$ 
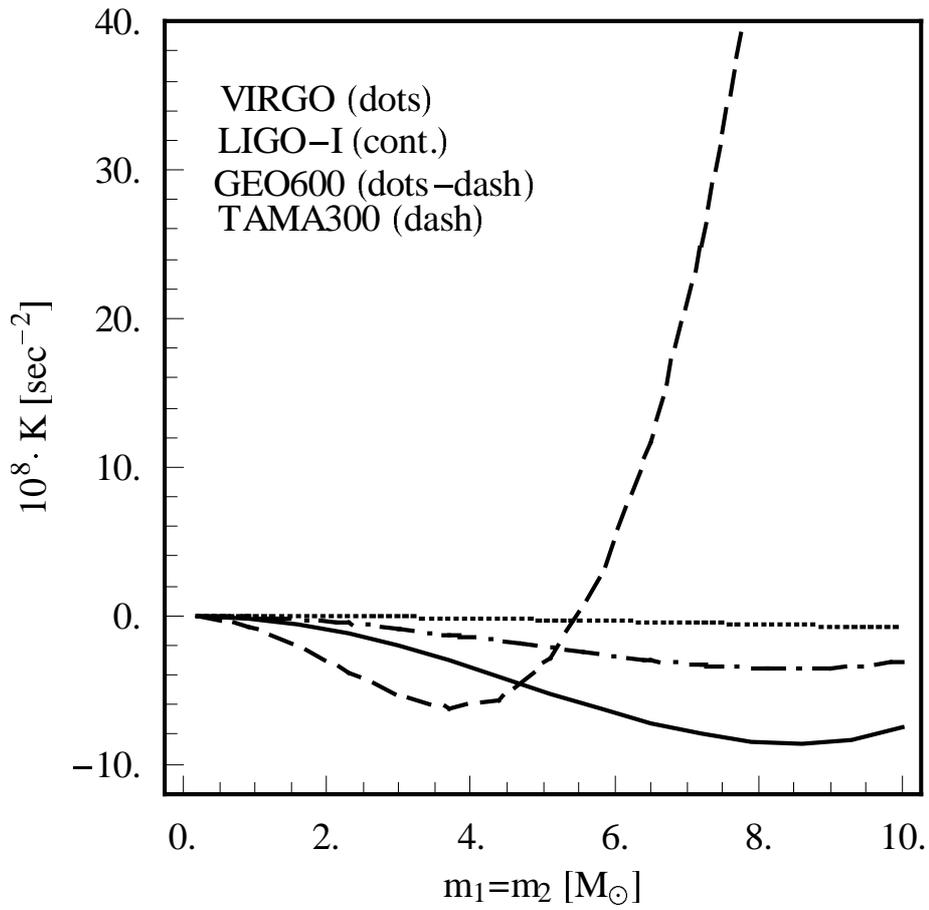
Fig. 2 - The gaussian curvature $K$ of ${\cal P}$ vs. $m_1\!=\!m_2$  
(worst case) for TAMA300, GEO600, LIGO-I and VIRGO.
$$~$$ 
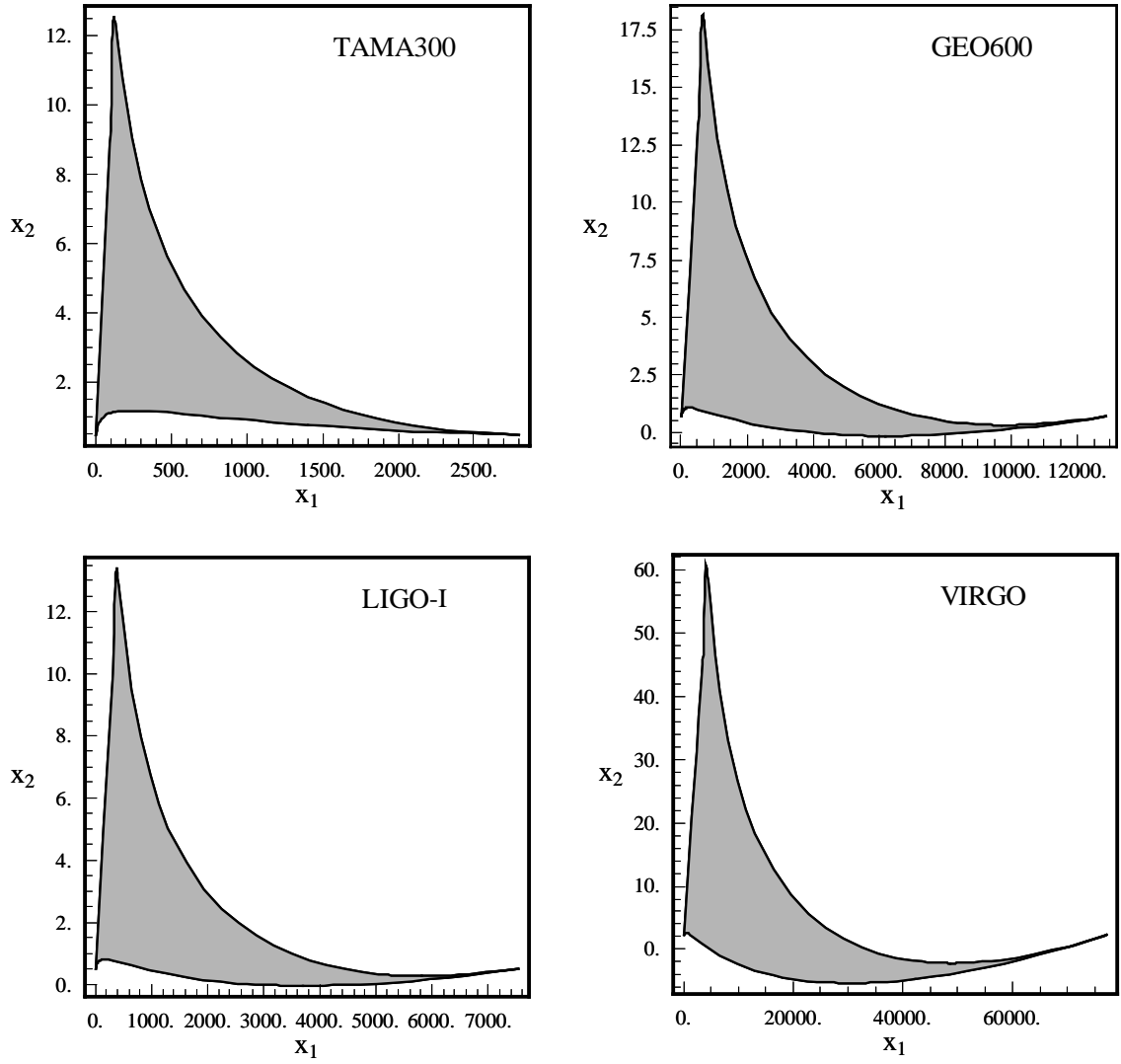
Fig. 3 - The flat simplexes ${\cal T}$ corresponding to
$0.2 M_{\odot} \leq m_1 \leq m_2 \leq 10 M_{\odot}$ for 
TAMA300, GEO600, LIGO-I and VIRGO.
$$~$$ 
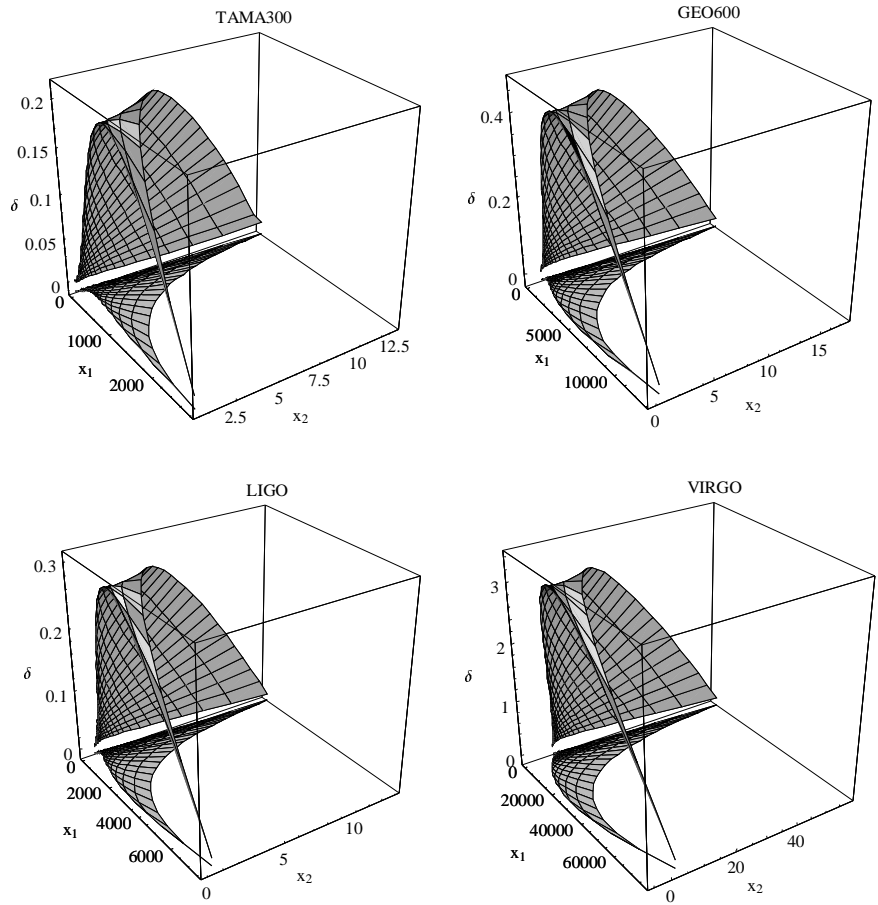
Fig. 4 - The euclidean distance $\delta=(x_3^2+x_4^2+x_5^2)^{1/2}$ between the manifolds ${\cal P}$ and ${\cal T}$ as a function of $x_1,~x_2$ for TAMA300, GEO600, LIGO-I and VIRGO.
$$~$$ 
Fig. 5 - The quantity $\eta$,  eq. (\ref{eq:bound}),  vs. $m_1$, $m_2$  for LIGO-I.
$$~$$ 
Fig. 6 - The quantity $\eta$,  eq. (\ref{eq:bound}),  vs. $m_1\!=\!m_2$  (worst case)  for TAMA300, GEO600, LIGO-I and VIRGO. 
$$~$$ 
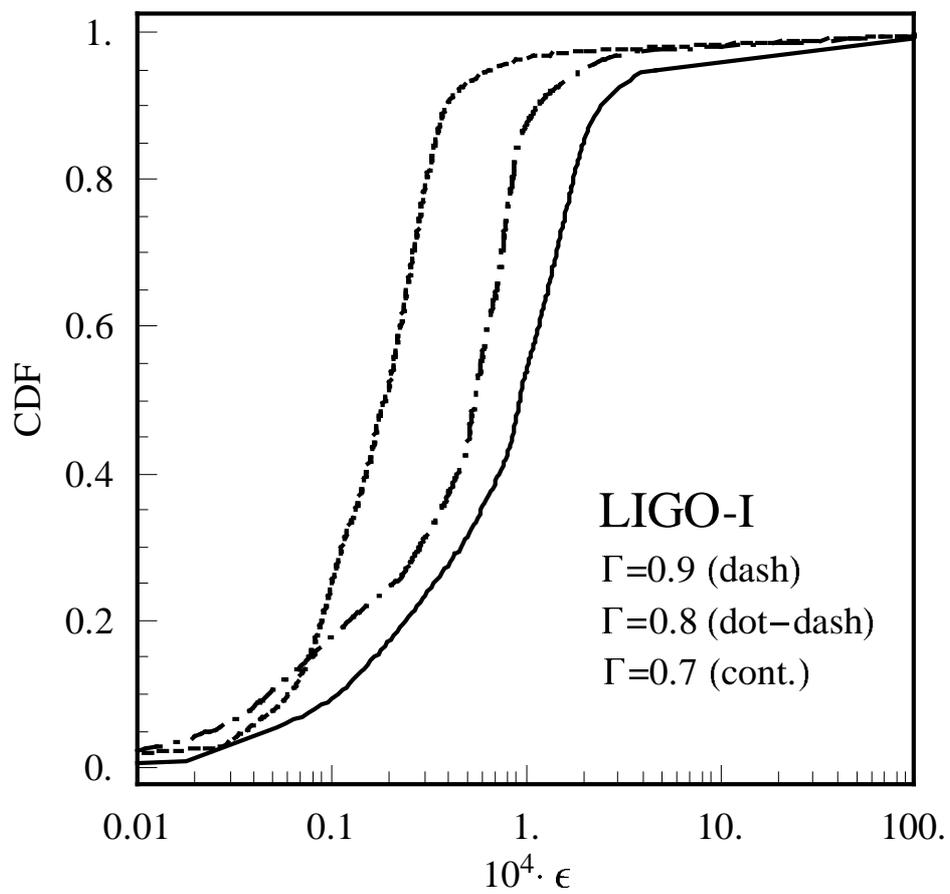
Fig. 7 - Cumulative distribution of $\epsilon$,  eq. (\ref{eq:epsilon}), 
for  LIGO-I at $\Gamma=0.7,~0.8,~0.9$.
$10^4$ trial sources in $0.2 M_{\odot} \leq m_1 \leq m_2 \leq 10 M_{\odot}$.  
$$~$$ 
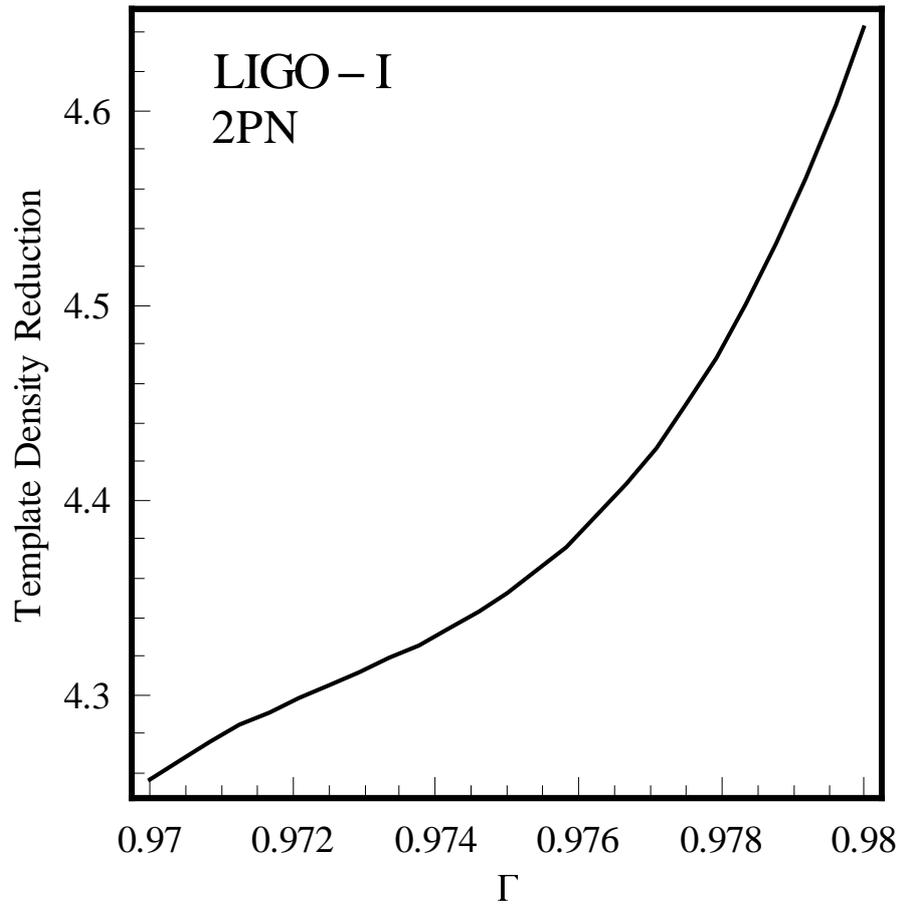
Fig. 8 - Template density reduction vs. $\Gamma$ after cardinal interpolation (2PN order templates, LIGO-I). 
$$~$$

\begin{thebibliography}{99}
\bibitem{3PN_stall}{PN  binary inspiral waveforms have been computed 
up to 2.5PN order.  Efforts to include 3PN terms have not yet
been completed (see e.g., L. Blanchet and G. Faye, gr-qc0007051).}
\bibitem{notation}{Our notation differs from that used in \cite{TaTa}. 
For better readability, we include explicitly the null 0.5PN term 
in the PN phase expansion (\ref{eq:phase})}.
\bibitem{parabolic}{The quadratic approximation (\ref{eq:match1}) 
is accurate in a neighbourhood of $M=1$, e.g., for LIGO-I,
for $M \gsim 0.95$.}
\bibitem{invar}{The spin-free phase is invariant 
upon exchange between $m_1$ and $m_2$.}
\bibitem{GaussK}{The global flatness of a 2D riemannian  manifold $({\cal P}, g)$
is gauged by its gaussian curvature  $K$ \cite{Spivak}:
$$
\vspace*{-.1cm}
K = \frac{R_{1212}}{det[g]},
$$
where the (only component of the) Riemann tensor is:
$$
\vspace*{-.3cm}
R_{1212}=\frac{1}{2}
\left(2 g_{12,12}-g_{11,22}-g_{22,11} \right)
+\sum_{m,n}^{1,2}
g^{mn}
\left\{  
\left[\mbox{\small{21}},m \right] \left[\mbox{\small{12}},n \right] - 
\left[\mbox{\small{11}},m \right] \left[\mbox{\small{22}},n \right] 
\right\}.
$$
In the above formula, for the case of our interest, $g_{pq}$ is 
given by (\ref{eq:2Dmetric}), and is a function of $m_1, m_2$; commas
denote partial derivatives, and 
$$
\left[ pq, r\right] = g_{pr,q}+g_{qr,p}-g_{pq,r}
$$
are the Christoffel symbols of the first kind. It is well known
that the 2D manifold ${\cal P}$ is flat, i.e., applicable to the plane,
iff  $K=0$.}
\bibitem{detail}{The partial derivatives in eq. (\ref{eq:gamma})
are readily computed from the known partial
derivatives of $\theta^i$ w.r.t. $m_j$, and the
inverse of the Jacobian matrix obtained from the
differential counterpart of eq. (\ref{eq:sys}).}
\bibitem{medians}{The median lines here are (more or less arbitrarily)
defined as those of the straight-edge triangle in $(x_1,x_2)$
whose vertices coincide with those of ${\cal T}$.}
\bibitem{project}{Equation $(A8)$ is readily obtained \cite{Owe96}
by projecting the metric on the subspace orthogonal to $\Delta\theta^0$. }
\end{thebibliography}
\end{document}